\documentclass[pre,onecolumn,preprint,superscriptaddress,showkeys,showpacs]{revtex4}
\usepackage{amssymb}
\usepackage{graphicx}
\usepackage[english]{babel}
\begin{document}

%%%%%%%%%%%%%%%%%%%%%%%%%%%%%%%%%%%%%%%%%%%%%%%%%%%%%%%%%%%%%%%%%%%%%%%%%%%%%%%%%%%%%%%%%%%%%%%%%%%%%%%%%%%%%%%%
\title{Nonextensive Interpretation Of Radiative Recombination In Electron Cooling}

\author{G.~Maero}
\email{g.maero@gsi.de}
\affiliation{Gesellschaft f\"{u}r Schwerionenforschung (GSI), D-64291, Darmstadt, Germany}
\author{P.~Quarati}
\email{piero.quarati@polito.it}
\affiliation{Dipartimento di Fisica, Politecnico di Torino, I-10129 Torino, Italy}
\affiliation{INFN - Sezione di Cagliari, I-09042 Monserrato, Italy}
\author{F.~Ferro}
\email{fabrizio.ferro@polito.it}
\affiliation{Dipartimento di Fisica, Politecnico di Torino, I-10129 Torino, Italy}
\affiliation{INFN - Sezione di Torino, I-10125 Torino, Italy}

\begin{abstract}
An interest for the low-energy range of the nonextensive distribution function arises from the study of radiative recombination in electron cooling devices in particle accelerators, whose experimentally measured reaction rates are much above the theoretical prediction. The use of generalized distributions, that differ from the Maxwellian in the low energy part (due to subdiffusion between electron and ion bunches), may account for the observed rate enhancement. In this work, we consider the isotropic distribution function and we propose a possible experiment for verifying the existence of a cut-off in the generalized momentum distribution, by measuring the spectrum of the X-rays emitted from radiative recombination reactions.
\end{abstract}

\date{\today}


\maketitle

%%%%%%%%%%%%%%%%%%%%%%
\section{Introduction}
%%%%%%%%%%%%%%%%%%%%%%
In the field of accelerator physics the request for cold beams, with low energy spread and emittance, is an important issue. The electron cooling technique, suggested by Budker in 1966, has been largely used  for this purpose~\cite{Parkhomchuk}. Nevertheless, the coexistence of positive ions and electrons leads to a certain amount of recombinations, and the study of this phenomenon has encountered a problem that is not completely solved up to now, i.e. the radiative recombination enhancement at low relative energy.

On the other hand, the nonextensive statistical mechanics has been successfully applied so far to various fields of physics, for instance in the study of astrophysical plasmas. Here we show the possible application of generalized statistics to radiative recombination; the results obtained go indeed into the right direction to account for the measured enhancement.

In the following sections we briefly summarize the electron cooling technique and radiative recombination, then propose the use of a superextensive distribution instead of a Maxwellian function in the calculation of the recombination rate because of the presence of correlations and/or random and memory effects among the particles, and finally we present theoretical rates from an isotropic distribution. Justifications and suggestions for experimental tests are also given to confirm the existence of a cut-off in the momentum distribution which, among all the generalized functions recently proposed, appears only in the Tsallis distribution.

%%%%%%%%%%%%%%%%%%%%%%%%%%%%%%%%%%%
\section{Cooling and recombination}
%%%%%%%%%%%%%%%%%%%%%%%%%%%%%%%%%%%
Liouville's theorem forbids the reduction of beam emittance by use of optical methods; therefore, external dissipative forces must be applied in order to overcome this huge problem in devices like a storage ring, where the high number of cycles would result in a heavy beam quality loss. In 1966, Budker suggested to overlap the beam of interest with a colder electron one, and to exploit Coulomb scattering in order to carry away the energy excess in the hot ion component. Shortly afterwards, devices built for this purpose proved the efficiency of the method.

Although their cross sections are orders of magnitude smaller, various phenomena can occur aside of Coulomb elastic scattering, such as ion excitation, ionization and several atomic recombination mechanisms, among which lies the radiative recombination,
\begin{displaymath}
A^{Z+}+e^{-}\rightarrow A^{\left(Z-1\right)+}+h\nu\, ,
\end{displaymath}
where an electron goes from free to bound state, and the energy excess is released through X-rays. Several experiments with bare ions (three-body processes like dielectronic recombination are forbidden) show an unexpected feature: when the relative velocity of electron-ion bunches approaches zero, the radiative recombination rate increases more than theoretically predicted on the basis of the calculated cross section and anisotropic Maxwellian distribution~\cite{Hoffknecht}.

The cross section of radiative recombination to any hydrogenic level with principal quantum number $n$ reads
\begin{equation}
\sigma_{n}^{RR}(E_{k})=2.10\cdot10^{-22}\frac{Z^{4}E_{1s}^{2}}{nE_{k}\left(Z^{2}E_{1s}+n^{2}E_{k}\right)}\,[\mathrm{cm}^2]\, ,\label{cross section n}
\end{equation}
where $E_{1s}$ is the Rydberg energy, $E_{k}$ is the electron kinetic energy in the electron-ion center-of-mass frame and $Z$ is the nuclear charge of the atom. A further quantum mechanical correction may be introduced via the so-called {\it Gaunt factor} $g_{n}\left(E_{k}\right)$, that is convenient as there are parametrizations or tables for its values.

The total cross section is obtained by summing the function in Eq.~\ref{cross section n} over all the accessible values of $n$ (often a proper upper limit $n_{max}$ is chosen); furthermore, a suitable parametrization of the total cross section yields
\begin{equation}
\sigma_{tot}\left(\varepsilon\right)=\sigma_{1s}\left(\varepsilon\right)s\left(\varepsilon\right)\, ,\label{total cross section}
\end{equation}
where $\varepsilon=\frac{E_{k}}{E_{1s}}$ and $s\left(\varepsilon\right)$ is a function of $\varepsilon$ evaluated numerically~\cite{Erdas}.

The recombination rate is then defined as an average of the cross section in Eq.~\ref{total cross section} either over the distribution functions of the two beams, or a distribution function of their relative velocity and, if we consider an already cooled ion beam, we can reduce ourselves to the average over the electron distribution $f({\bf v})$ alone; thus, the definition becomes
\begin{displaymath}
\alpha^{RR}=\left\langle v\sigma_{tot}\left({\bf v}\right)\right\rangle =\int\mathrm{d}^{3}{\bf v}\, v\sigma_{tot}\left({\bf v}\right) f\left({\bf v}\right)\, .
\end{displaymath}

Usually, an anisotropic Maxwellian distribution is considered: in the transverse plane the function is isotropic, whereas longitudinally the width (temperature) of the distribution is different. The distribution then reads
\begin{displaymath}
f\left({\bf v}\right)=\frac{m}{2\pi kT_{\bot}}\left(\frac{m}{2\pi kT_{\Vert}}\right)^{\frac{1}{2}} \exp\left[-\frac{1}{2}\frac{m\left(v_{\Vert}-v_{rel}\right)^{2}}{kT_{\Vert}}\right]\exp\left[-\frac{1}{2}\frac{mv_{\bot}^{2}}{kT_{\bot}}\right]\, ,
\end{displaymath}
where $v_{rel}$ ($E_{rel}=\frac{1}{2}mv_{rel}^2$) is the relative velocity (energy) of the two bunches of particles (electrons and ions).

The two temperatures $T_{\Vert}$ and $T_{\bot}$ are the typical parameters defining the experimental conditions; their values are around tens of meV for $T_{\bot}$ (strongly determined by the voltage at which the thermocathode that produces the electron cloud operates) and two or three orders of magnitude lower for $T_{\Vert}$. 

A clear example of recombination enhancement (showing a value up to $5.2$ at zero relative energy) is the experiment with $Bi^{83+}$ ions, performed in the ESR at GSI~\cite{HoffknechtArxiv}.

%%%%%%%%%%%%%%%%%%%%%%%%%%%%%%%%%%%%%%%%%%%%%%%%%%%%%%%%%%%%%%%%%%%%%%%%%%%%%%%%%%%%%%%%%%%%%%%%
\begin{figure}[t]

\begin{center}
\includegraphics[width=0.5\textwidth]{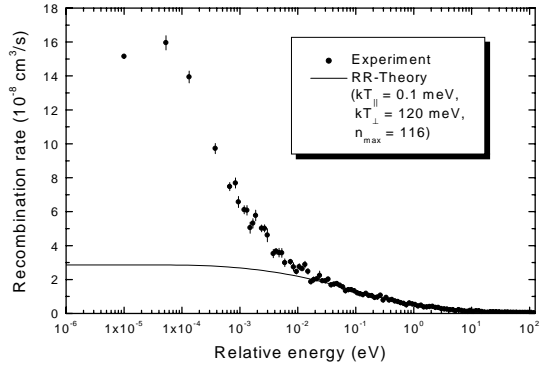}
\end{center}

\caption{Comparison between theory (Maxwellian rate) and experimental data in the case of the $Bi^{83+}$ experiment performed at GSI (picture taken from Ref.~\cite{HoffknechtArxiv}).}\label{enhancement figure}

\end{figure}
%%%%%%%%%%%%%%%%%%%%%%%%%%%%%%%%%%%%%%%%%%%%%%%%%%%%%%%%%%%%%%%%%%%%%%%%%%%%%%%%%%%%%%%%%%%%%%%%

There are various interpretations of the enhancement: for the authors of Ref.~\cite{Gwinner}, energy, due to the presence of a magnetic field, is transferred from the longitudinal to the transverse motion and therefore there is an increase in the time spent by the electron in proximity to the ion, enhancing the recombination probability.

Furthermore, a work by Heerlein, Zwicknagel and Toepffer~\cite{Heerlein} takes into account the effect of the external fields, considered as the cause of enhancement, suggesting that this arises from a field-driven process, i.e. electrons that acquire negative energy upon the merging of the electron and ion beams, because of the presence of the toroidal magnetic field that permits the bending of the cooling electrons onto the trajectory of the ion beam; the magnet separates the loosely bound electrons of the most outer atomic levels, so that these can yield an additional recombination channel; hence, we have to count not only free-bound transitions, but also bound-bound transitions. See Ref.~\cite{Hoerndl} for criticisms to this explanation.

So far, no complete theory has been proposed, anyway, and scaling laws for the enhancement require fitting upon experimental data; on the contrary, we suggest an alternative approach based on a self-consistent thermostatistical framework, namely the nonextensive statistical mechanics, where the key idea is the use of a deformed distribution function.

%%%%%%%%%%%%%%%%%%%%%%%%%%%%%%%%%%%%%%%%%%%%%%%%%
\section{Non-Maxwellian approach: isotropic case}
%%%%%%%%%%%%%%%%%%%%%%%%%%%%%%%%%%%%%%%%%%%%%%%%%
The nonextensive statistical mechanics, introduced by Tsallis in the last two decades, can be seen as a generalization of Maxwell-Boltzmann statistics; the Tsallis distribution function reads
\begin{displaymath}
f_{q}\left(v\right)=B_{q}\left[1-\left(1-q\right)\frac{mv^{2}}{k_{B}T_{q}}\right]^{\frac{1}{1-q}}\, ,
\end{displaymath}
where $q$ is the nonextensivity index, $B_{q}$ is the normalization factor and $T_{q}=\frac{2}{5-3q}T_{MB}$ is the $q$-temperature, that must be used every time we want to relate Boltzmann and Tsallis statistics~\cite{Tsallis}.

The distinctive feature of the $q$-distribution is that when $q>1$ (subextensivity and superdiffusion), $f_{q}$ shows a lower peak at $v=0$ with respect to the Maxwellian function $f_{MB}$, while the tail at higher velocities is increased; for $q<1$ (superextensivity and subdiffusion), the situation is the other way round: $f_{q}$ is higher-peaked and narrower, so that the high-velocity tail is accordingly depleted, and a cut-off appears at $E_{cut}=\frac{k_{B}T}{1-q}$.

Since the radiative recombination cross section shows a peak for $v\rightarrow 0$, while the particles in the tail have scarcely an influence, a superextensive distribution, increasing the number of particles around the zero-energy point, leads to an increase in $\alpha^{RR}$. This is calculated according to the following formula:
\begin{equation}
\alpha_{q}^{RR}=\left\langle\left\langle v\sigma_{tot}\left(v\right)\right\rangle\right\rangle_{q}=\frac{\int \mathrm{d}^{3}{\bf v} f_{q}^{q}\left(v\right)v\sigma_{tot}\left(v\right)}{\int \mathrm{d}^{3}{\bf v} f^{q}\left(v\right)}\, .\label{non-extensive rate}
\end{equation}

Expressing velocity $v$ as function of the kinetic energy $E_{k}$, and using some analytical properties, we obtain from Eq.~\ref{non-extensive rate},
\begin{displaymath}
\alpha_{q}^{RR}=B_{q}8\pi\frac{1}{\left(Mc^{2}\right)^{2}}c^{4}\int\mathrm{d}E_{k}\left[1-\left(1-q\right)
\frac{E_{k}}{kT_{q}}\right]^{\frac{q}{\left(1-q\right)}}\sigma_{tot}\left(E_{k}\right)E_{k}\, ,
\end{displaymath}
where the integration limits are $\left[0,\infty\right)$ for $q>1$, or $\left[0,E_{cut}\right)$ for $q<1$.

\begin{figure}[t]

\begin{center}
\includegraphics[width=0.5\textwidth]{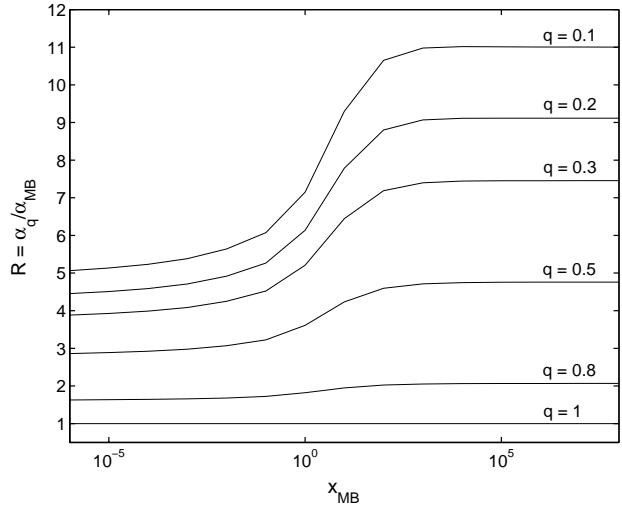}
\end{center}

\caption{Plot of the recombination enhancement $R$ against the dimensionless temperature $x=kT_{MB}/E_{1s}$ for different values of the $q$ parameter in the superextensivity interval, ranging from $q=0.1$ to $q=1$. (Calculations performed for an electron-positron system).}\label{ratio figure}

\end{figure}

Two important remarks: first, we adopted an isotropic distribution function because the $q$-distribution cannot be rigorously factorized; therefore, coherently, we shall compare it with an $\alpha_{MB}^{RR}$ obtained as a result of an isotropic Maxwellian. Inserting the parametrization of Ref.~\cite{Erdas}, we get the recombination enhancement at zero relative energy, defined as the ratio $R=\frac{\alpha_{q}}{\alpha_{MB}}$. In Fig.~\ref{ratio figure} we plot the ratio $R$ as a function of the dimensionless temperature $x=\frac{kT_{MB}}{E_{1s}}$ in the case of an electron-positron system. The curve shows that actually, as $q\rightarrow 0$, a recombination enhancement appears.

Second, our calculations show that the recombination rate is almost the same for both Maxwellian and Tsallis distribution functions at the same temperature; the major influence on the enhancement is therefore the temperature alteration given by $q$ in the $T_q$ definition.

%%%%%%%%%%%%%%%%%%%%%%%%%%%%%%%%%%%%%%%%%%%%%%%%%%%%%%%%%%%%%%%%%%%%%%%%%%%%%%%%%%%%%%%%%%%%%%%%
%%%%%%%%%%%%%%%%%%%%%%%%%%%%%%%%%%%%%%%%%%%%%%%%%%%%%%%%%%%%%%%%%%%%%%%%%%%%%%%%%%%%%%%%%%%%%%%%

%%%%%%%%%%%%%%%%%%%%%%%%%%%%%%%%%%%%%%%%%%%%%%%%%%%%%%%%%
\section{Justifications, testing and gedanken Experiment}
%%%%%%%%%%%%%%%%%%%%%%%%%%%%%%%%%%%%%%%%%%%%%%%%%%%%%%%%%
Many studies demonstrated that there exists a set of non-Maxwellian distribution functions that are stationary solutions of dynamical equations~\cite{Kaniadakis}. For example, when a statistical system of particles is subjected to external random forces or electric fields (condition that also applies to radiative recombination in electron cooling devices) the actual functional dependence on energy of its distribution function relies strongly on the microscopical interaction acting among the ions. It was shown, in particular, that an interaction cross section roughly constant with energy naturally leads to the Tsallis distribution that was used throughout this paper~\cite{Ferro}.

Moreover, the presence of a cut-off in a distribution function is not a completely new feature: already in 1940, Spitzer imposed a cut-off to the Maxwellian distribution in order to explain the stability of galaxy clusters despite the escape velocity effect~\cite{Spitzer}. Furthermore, a theoretical explanation of the cut-off was given by relating it to a finite heat reservoir~\cite{Plastino}.

Many experimental results show how the rate coefficient behaves as a function of $E_{rel}$. Among others, $C^{6+}$, $F^{6+}$, $Ne^{10+}$ and $Bi^{83+}$ ion beams have been used~\cite{Hoffknecht,Gwinner,Gao}. All experiments indicate that, respect to $E_{rel}$, the rate coefficient decreases much faster than the Maxwellian one calculated at a temperature $T_{q}$ lower than the Maxwellian $T_{MB}$, in spite of the fact that at $E_{rel}=0$ the Maxwellian rate at $T_{q}$ satisfactorily approximates the experimental result. Only imposing a cut-off in the momentum distribution we can reproduce, for $E_{rel}\neq 0$, the experimental behaviour (see Fig.~\ref{enhancement figure}). The previous argument actually represents an effective test that a cut-off naturally exists in the momentum distribution.

A further possible test to verify the presence of a superextensive distribution would be the observation of the recombination X-ray spectrum. A measured cut-off energy $E_{cut}$, beyond which there are no electrons, would indeed imply the presence of a corresponding cut-off in the energy spectrum of the radiation emitted from recombination: the maximum observable energy might thus be
\begin{displaymath}
E_{spectrum}^{max}=E_{cut}+E_{1s}\, ,
\end{displaymath}
since $E_{1s}$ is the largest binding energy. See Ref.~\cite{Poth} for a detailed description of the X-ray spectrum measurements, which could be possible in the future. But, since this kind of measurements are out of the present experimental efficiency, this remains a gedanken experiment.

%%%%%%%%%%%%%%%%%%%%%
\section{Conclusions}
%%%%%%%%%%%%%%%%%%%%%
In order to explain the atomic recombination rate coefficient measured in electron cooling devices in storage rings (of a factor up to 5 greater than its Maxwellian evaluation) we must consider that correlations and memory effects in addition to dissipative and random effects in the electron gas impose, for its description, the use of a deformed momentum distribution function. The superextensive Tsallis distribution seems to be appropriate, among many other different deformed distributions proposed, because, among other reasons, this distribution contains an energy cut-off we would expect as a consequence of our previous considerations.

The presence of cut-off results by examining the experimental measurements of the rate at relative energies different from the very low relative energy region. This fact is a test of the validity, in this particular situation, of the Tsallis nonextensive thermostatistics.

\end{document}